\documentclass[aoas,MSNbibl,nameyear,seceqn,dvips]{arximspdf}
\usepackage{dcolumn}
\usepackage{graphicx}

\doi{10.1214/11-AOAS512}
\volume{6}
\issue{1}
\pubyear{2012}
\firstpage{161}
\lastpage{185}

\makeatletter

\newcolumntype{d}[1]{D{.}{.}{#1}}

\newcommand{\N}{\mathrm{N}}
\newcommand{\U}{\operatorname{Unif}}
\newcommand{\C}{\mathrm{C}}
\newcommand{\Be}{\operatorname{Be}}
\newcommand{\dd}{\mathrm{d}}
\newcommand{\HIB}{\operatorname{HIB}}
\newcommand{\HB}{\operatorname{HB}}
\newcommand{\E}{\mathrm{E}}
\newcommand{\Var}{\operatorname{Var}}
\newcommand{\p}{p}

\newcommand{\bbeta}{\bolds{\beta}}
\newcommand{\by}{\mathbf{y}}

\newcommand{\kummer}{{}_1 F_1}
\newcommand{\kummertwo}{{}_2 F_1}

\newtheorem{theorem}{Theorem}[section]

\makeatother

\begin{document}
\begin{frontmatter}

\title{Good, great, or lucky? Screening for firms with sustained
superior performance using heavy-tailed~priors}
\runtitle{Screening for firms with superior performance}

\begin{aug}
\author[A]{\fnms{Nicholas G.} \snm{Polson}\ead[label=e1]{ngp@chicagobooth.edu}}
\and
\author[B]{\fnms{James G.} \snm{Scott}\corref{}\ead[label=e2]{james.scott@mccombs.utexas.edu}}
\runauthor{N. G. Polson and J. G. Scott}
\affiliation{University of Chicago and University of Texas at Austin}
\address[A]{Booth School of Business\\
University of Chicago\\
5807 South Woodlawn Avenue \\
Chicago, Illinois 60637-1610 \\
USA\\
\printead{e1}}
\address[B]{Division of Statistics\\
\quad and Scientific Computing\\
University of Texas at Austin\\
1 University Station, B6500\\
Austin, Texas 78712\\
USA\\
\printead{e2}} %adresu isvedimo komanda gale!
\end{aug}

% HISTORY:
\received{\smonth{10} \syear{2010}}
\revised{\smonth{9} \syear{2011}}

% ABSTRACT
%
\begin{abstract}
This paper examines historical patterns of ROA (return on assets) for a
cohort of 53,038 publicly traded firms across 93 countries, measured
over the past 45 years. Our goal is to screen for firms whose ROA
trajectories suggest that they have systematically outperformed their
peer groups over time. Such a project faces at least three statistical
difficulties: adjustment for relevant covariates, massive multiplicity,
and longitudinal dependence. We conclude that, once these difficulties
are taken into account, demonstrably superior performance appears to be
quite rare. We compare our findings with other recent management
studies on the same subject, and with the popular literature on
corporate success.

Our methodological contribution is to propose a new class of priors for
use in large-scale simultaneous testing. These priors are based on the
hypergeometric inverted-beta family, and have two main attractive
features: heavy tails and computational tractability. The family is a
four-parameter generalization of the normal/inverted-beta prior, and is
the natural conjugate prior for shrinkage coefficients in a
hierarchical normal model. Our results emphasize the usefulness of
these heavy-tailed priors in large multiple-testing problems, as they
have a~mild rate of tail decay in the marginal likelihood
$m(y)$---a~property long recognized to be important in testing.
\end{abstract}

% KEYWORDS
%
\begin{keyword}
\kwd{Corporate benchmarking}
\kwd{type-II beta distribution}
\kwd{multiple testing}
\kwd{normal scale mixtures}
\kwd{sparsity}.
\end{keyword}

\end{frontmatter}

%s1 #&#
\section{Introduction}

%s1.1 #&#
\subsection{Large-scale screening of historical ROA data}
Understanding the reasons why some firms thrive and others fail is one
of the primary goals of research in strategic management. Studies that
examine successful companies to uncover the putative secrets of
successful companies are very popular, both in the academic and popular
literature.

\begin{figure}

\includegraphics{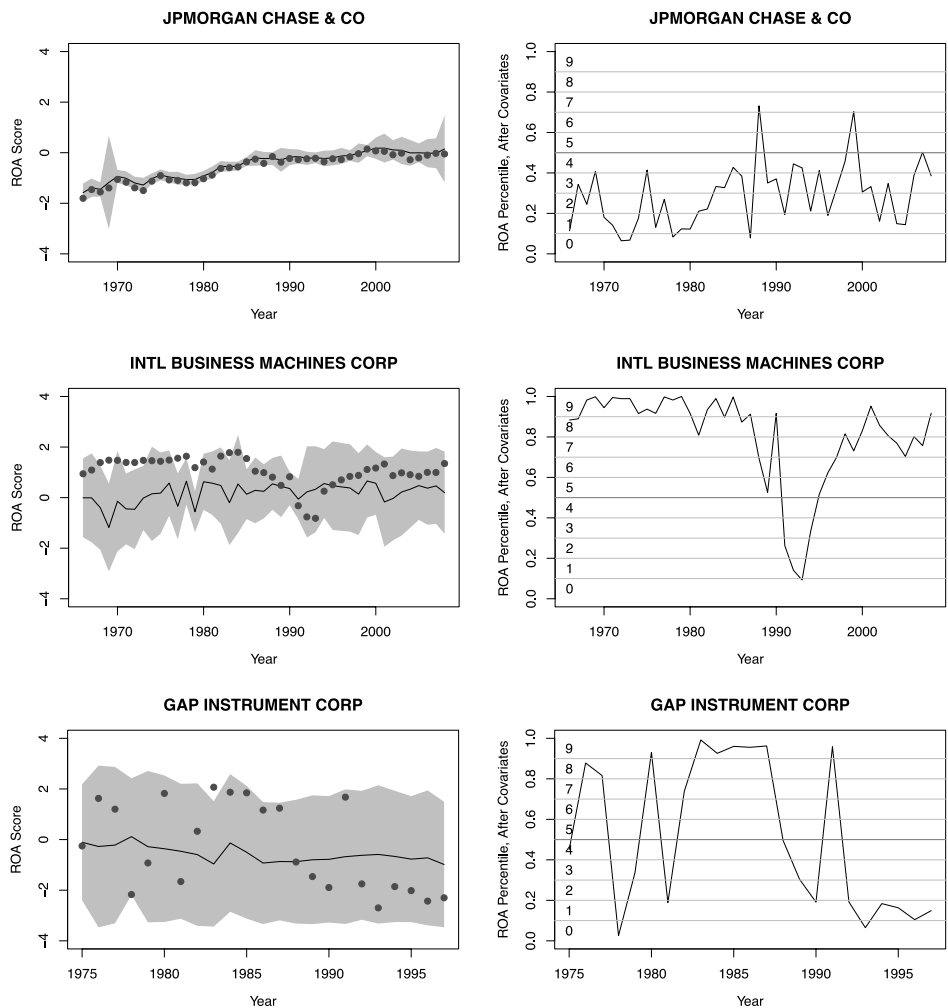}

\caption{Left: the actual performance of three firms (dots),
superimposed on the benchmark distribution estimated from the Bayesian
regression-tree model (black line and grey area, showing the posterior
mean and $95\%$ predictive interval of expected performance by all
firms in the corresponding peer group). Right: these same firms placed
on a common (normal CDF) scale of benchmarked performance, with the
integers 0--9 representing the decile.}
\label{examples1}
\vspace*{6pt}
\end{figure}

Before the search for special causes can begin, however, success must
be quantified and benchmarked. This is what our paper tries to do. In
keeping with prior studies
[\citet{mcgahanporter1999}, \citet{wigginsruefli2005},
\citet{hendersonetal2009}], we use a
common metric called ROA, or return on assets, to measure a company's
success. This quantity gives investors some notion of how effectively a
firm uses its available funds to produce income. It is fundamentally
different from a market-based measure like stock returns, which may
fail to reflect underlying fundamentals over long periods of time
(e.g., during bubbles), and which exhibit wild fluctuations that make
the identification of trends problematic. Figure \ref{examples1} shows
three examples of firm-level ROA trajectories over time; these have
been standardized using a procedure which we will soon describe.

%
%f1 #&#
%

In this paper we apply Bayesian methods to historical ROA data, with
the goal of comparing publicly traded companies against their peers. To
be sure, ROA is an imperfect measure of corporate success, and our
study will have the same shortcomings in this regard as any other that
uses ROA as an outcome variable. One important practical reason for our
use of ROA, aside from a desire to use the same metric as other
researchers studying similar questions, is the sheer availability of
data on companies from across the world (rather than just in the United
States). This enables us to screen as large a database as possible:
645,456 records from 53,038 companies in 93 different countries,
spanning 1966--2008. In principle, however, our Bayesian statistical
methodology could be applied to any outcome variable in any
subpopulation of the corporate universe.

We conclude that evidence of sustained superior performance is quite
rare. To reach this conclusion, we use Bayesian models to compute the
posterior probability that a firm falls into each of two classes: a
null class, wherein deviations from the peer-group average are
attributable to chance; and an alternative class, wherein these
deviations, both positive and negative, are systematic. These posterior
probabilities depend upon the particular assumptions made about the
longitudinal persistence of ``lucky'' performances, in a manner soon to
be explained. But even under the generous (and unrealistic) assumption
of longitudinal independence, we find that there are at most 1076 firms
over the last 45 years for which there is moderately strong evidence of
sustained superior performance over 5 years or more. We argue that this
is a conservative upper bound on the number of such firms, and that the
actual number is much smaller---our best estimate is 262, or $0.5\%$ of
all firms, once longitudinal dependence is taken into account.

%s1.2 #&#
\subsection{Statistical issues in identifying sustained superior performance}

Any attempt to benchmark performance, and to identify sustained
superior performers, must deal with at least three statistical challenges.

First, one must adjust observed performance for relevant covariates.
One important covariate is a firm's country of operation. Another one
is a firm's industry; as \citet{hendersonetal2009} observe, some
industries exhibit structures that are intrinsically more favorable to
monopolies, which would seem to be a source of advantage unrelated to
managerial talent or firm-level characteristics. Other potentially
important characteristics that have been explored in the literature
include a firm's size and capital structure.

Our method adjusts for the effect of all of these covariates, both on
the conditional mean and conditional variance of performance.
Importantly, there is no reason to assume that ROA depends upon them
linearly. This is quite different from the situation in finance, for
example, where the capital-asset pricing model (CAPM) and its variants
predict a linear dependence between firm-level and market-level
measures of performance. No such theory exists that would predict a
parallel result for ROA. This means that nonlinear relationships must,
at least in principle, be allowed. We do this using Bayesian
treed-regression models, as described in Section \ref{sec4}.

Second, even ``lucky'' performance trajectories may exhibit significant
longitudinal dependencies that lead to spurious declarations of
impressiveness. Following \citet{denrell2005}, imagine a very simple
state-space model, wherein
\begin{eqnarray*}
y_{t} &=& a x_t + e_t, \\
x_t &=& b x_{t-1} + u_t ,
\end{eqnarray*}
where $y_t$ is an observed performance metric, and $x_t$ is some
underlying AR(1) firm-level characteristic (e.g., resources). Even if
there is no systematic component of variation in $x_t$, the observed
$y_t$'s can still exhibit pronounced longitudinal autocorrelation,
which can look very much like a sustained run of excellence. Formally
correcting for such autocorrelation would require specific parametric
models incorporating a wide variety of firm-level effects. Instead of
taking this route, we try to correct for longitudinal dependence in a
crude-but-simple fashion by estimating an effective sample size for
each firm, and adjusting our Bayesian model accordingly.

Finally, there is the issue of massive multiplicity. Given the large
number of hypothesis tests being conducted, and the frequentist
leanings of the management-theory community, maintaining control over
false positives is crucial. Yet having access to the posterior
distribution of effect sizes can greatly inform follow-up case studies
of individual firms, and is only possible under a fully Bayesian model.
This applied context makes a combined Bayes/frequentist approach
especially appealing.

Our paper's methodological innovation is to introduce a new class of
heavy-tailed priors for the multiple-testing problem. We first give a
brief overview of this problem from a Bayesian perspective (Section \ref{sec2}),
deferring much of the details to \hyperref[app]{Appendices}. We then
describe some
simulation studies in Section \ref{sec3}, which are designed to
benchmark our
proposed method against reasonable alternatives. In these studies, our
methods show excellent performance in terms of limiting false
positives, lending credence to the results for the actual data.
Finally, we analyze the corporate ROA data in Section \ref{sec4}, where
we also
describe in further detail how we approach the other statistical issues
we have raised.

%s2 #&#
\section{Large-scale simultaneous testing}\label{sec2}

%s2.1 #&#
\subsection{Methodological overview}

In large-scale simultaneous testing, the\break goal is to uncover
lower-dimensional signals from high-dimensional data. For example,
researchers who use microarrays have long been interested in the
problem of multiplicity adjustment, where ``adjustment'' can be
understood in the sense of adjusting one's tolerance for surprise as
the set of potentially surprising events grows large. The same issue
arises in all modern high-throughput experiments; other examples
include functional magnetic-resonance imaging, environmental sensor
networks, combinatorial chemistry, and proteomics. Too many type-I
errors will mean too many expensive wild-goose chases. Hence, the case
for a testing procedure that displays good frequentist properties is
very compelling.

But so too is the case for a model-based Bayesian procedure. These
experiments may involve thousands of separate tests, and such a large
volume of data often allows the distributional properties of
``signals'' and ``noise'' to be characterized quite precisely.

This paper considers a new version of the two-groups multiple-testing
model, where we observe data $y_i$ for $i = (1, \ldots, p)$ according
to a hierarchical model:
\begin{eqnarray*}
(y_i \mid\beta_i, \sigma^2) &\sim& \N(\beta_i, \sigma^2), \\
(\beta_i \mid w, \theta) &\sim& w \cdot g(\beta_i \mid\theta) +
(1-w) \cdot\delta_0, \\
w &\sim& p(w) ,
\end{eqnarray*}
a mixture of a Dirac measure at zero, and an alternative model $g$ that
is absolutely continuous with respect to the Lebesgue measure. (The
alternative model $g$ has hyperparameter $\theta$, presumably also
given a prior.) The most attractive feature of this model is that it
automatically adjusts for multiplicity, without the need for ad-hoc
regularization. This is because inference for the $\beta_i$'s will
involve the posterior for common mixing fraction, $p(w \mid\by)$. If
one tests many noise observations in the presence of a few signals,
then our estimate of $w$ will be small, making it more difficult for
all the observations to overcome the prior belief in their irrelevance.
This exerts a powerful form of control over false positives.

To handle the multiple-testing problem, we introduce a family of
distributions $g$ based on normal variance mixtures, where the mixing
distribution is a hypergeometric inverted-beta (HIB) prior:
\begin{eqnarray*}
(\beta_i \mid\lambda_i^2, \gamma_i = 1) &\sim& \N(0, \sigma^2
\lambda_i^2), \\
\lambda_i^2 &\sim& \HIB(a,b,\tau,s) ,
\end{eqnarray*}
where the indicator $\gamma_i = 1$ if $\beta_i$ is nonzero, and zero
otherwise. We approach these priors from a hybrid Bayesian/frequentist
perspective, using them to compute not only posterior distributions,
but also false-discovery rates, or FDR [\citet{benjamini1995}]. We also
study the behavior of the posterior mean, which is competitive with
existing gold-standard methods [e.g., \citet{johnstonesilverman2004}]
under squared-error loss.

In both our data analysis and simulation studies, we focus on three key
features of our approach:
\begin{longlist}[(1)]
\item[(1)] The hypergeometric inverted-beta scale mixtures form an
especially flexible class of symmetric, unimodal densities and can
accommodate a wide range of tail behavior and behavior near the
centering parameter. This class simultaneously generalizes the robust
priors of \citet{strawderman1971} and \citet
{BergerAnnals1980}, the
normal-exponential-gamma prior of \citet{griffinbrown2005}, and the
horseshoe prior of \citet{CarvalhoPolsonScott2008a}. The ability
of our
class to model heavy-tailed distributions with minimal computational
fuss is of particular relevance in testing problems [see, e.g.,
Section~5.2 of \citet{jeffreys1961}].
\item[(2)] Our class of priors allows very easy computation of a wide array
of important Bayesian and frequentist quantities. This includes
posterior means, variances, and higher-order moments; posterior null
probabilities for individual observations; the score function;
false-discovery rates; and local false-discovery rates
[\citet{efron2008}]. The ease with which these quantities can be
computed all
relates to the analytical tractability of the marginal likelihood
function $m(y)$, whose importance we describe in Section \ref
{SecMLFunction}. Appendix~\ref{appA} provides all the details.
\item[(3)] Our approach yields testing error rates that are competitive with
existing cutting-edge methods. At the same time, it also retains the
advantages of a fully Bayesian procedure, in that in principle one has
access to the joint posterior distribution of all parameters.
\end{longlist}

Many of the technical details characterizing the behavior of the basic
mixture model can be found in \citet{scottberger06} and
\citet{bogdanghosh2008b}. These authors assume that the nonzero
means follow a~normal distribution, an assumption we generalize in this paper.
\citet{domuller2005} also provide an interesting variation wherein the
nonzero means are modeled nonparametrically using Dirichlet processes.

The same issues arise in empirical-Bayes analysis. See, for example,
\citet{johnstonesilverman2004}, \citet{abramovichetal2006}
and \citet{dahlnewton2007}. Additionally, \citet{mulleretal2006},
\citet{bogdanetal2008} and \citet{parkghosh2010}. All describe the
relationship between Bayesian multiple testing and classical approaches
that control the false-discovery rate.

%s2.2 #&#
\subsection{The importance of the marginal likelihood function}
\label{SecMLFunction}

Many common Bayesian and frequentist treatments of the
multiple-testing\vadjust{\goodbreak}
problem can be understood through the marginal likelihood functions
\begin{eqnarray*}
m_0(y \mid\sigma^2) &=& \N(y \mid0, \sigma^2), \\
m_1(y \mid\theta) &=& \int_{\mathbb{R}} \N(y_i \mid\beta_i,
\sigma^2) g(\beta_i \mid\theta) \,\dd\beta_i, \\
m(y \mid\theta, \sigma^2) &=& w \cdot m_1(y) + (1-w) \cdot m_0(y) .
\end{eqnarray*}

First, following \citet{efron2008}, the local FDR and the posterior
probability of $y_i$ being noise are given by the same expression:
\[
fdr(y) = P(\beta_i = 0 \mid y, \sigma^2, \theta) = \frac{(1-w)
\cdot m_0(y)}{m(y)} .
\]
Furthermore, if we let $F_0(y) = \int_{-\infty}^y m_0(u) \,\dd u$,
$F_1(y) = \int_{-\infty}^y m_1(u) \,\dd u$, and $F(y) = w \cdot F_1(y)
+ (1-w) \cdot F_0(y)$, then the FDR is the tail area
\[
\operatorname{FDR}(y) = \frac{(1-w) \cdot F_0(y)}{F(y)} .
\]

Second, the marginal likelihood function also arises in Masreliez's
classic representation of the posterior mean. This gives an explicit
expression for the Bayes estimator for $\beta_i$ under squared-error
loss (assuming that $\gamma_i = 1$):
\[
\E(\beta_i \mid y, \gamma_i = 1) = y_i + \frac{d}{dy_i} \ln
m_1(y_i) ,
\]
versions of which appear in \citet{masreliez1975}, \citet{polson1991},
\citet{pericchismith1992} and \citet
{CarvalhoPolsonScott2008a}. The
choice of alternative model $g(\beta_i \mid\theta)$ is crucial,
insofar as it helps to determine $m_1(y)$.

At the same time, the prior should have desirable statistical
properties, with flat tails being a particularly important feature. The
use of heavy-tailed priors for constructing robust shrinkage estimators
has a long history, with prominent examples to be found in
\citet{strawderman1971} and \citet{BergerAnnals1980}.
Jeffreys, meanwhile,
observed as early as 1939 that heavy-tailed priors play an important
role in Bayesian hypothesis testing [see \citet{jeffreys1961}, a later
edition]. His arguments have been recapitulated in the context of
linear models by \citet{zellnersiow1980} and, more recently,
\citet{liangpaulo07}.

The difficulty is that, while heavy-tailed priors lead to a desirably
mild rate of tail decay in the marginal likelihood $m(\by)$, there are
few such priors that are also analytically tractable. Any prior that
possesses both properties, as our proposed family does under certain
hyperparameter choices, is therefore of great potential interest to
Bayesians and non-Bayesians alike.

We describe the hypergeometric-beta family of priors more fully in
a~leng\-thy technical \hyperref[app]{Appendix}. But first we present
simulation studies
that demonstrate the usefulness of our approach for limiting false
positives, before turning to an analysis of the data set at hand.

%s3 #&#
\section{Simulation studies}
\label{sec3}

As our methodological \hyperref[app]{Appendix} shows, hypergeometric
inverted-beta
scale mixtures of normals are an especially useful class of priors for
building discrete mixture models for $\beta_i$, due to the existence
of simple expressions for moments and marginals under the hypothesis
that $\beta_i$ is nonzero:
%
%e3.1 #&#
%e3.2 #&#
%
\begin{eqnarray}
(\beta_i \mid\kappa_i) &\sim& w \cdot\N(0, \kappa^{-1} - 1) +
(1-w) \cdot\delta_0, \\
\kappa_i &\sim& \HB(a,b,\tau, s) ,
\end{eqnarray}
where $\delta_0$ indicates a degenerate distribution at 0. The
posterior mean under this model is a natural estimator for $\bbeta=
(\beta_1, \ldots, \beta_p)$, since it averages over uncertainty
about whether each component is zero or nonzero.

We conducted two simulation studies comparing the mean-squared error
performance of our estimators with the procedure from
\citet{johnstonesilverman2004}, where $\beta_i$ is estimated by the
posterior median under a~mixture of a point mass zero and a
double-exponential (Laplace) prior. We also keep track of the number of
false positives generated by each procedure.

Each of the two studies involved estimating signals from a different
signal class. In all cases the dimension of the location vector was
$p=1\mbox{,}000$.
\begin{longlist}
\item[\textit{Experiment} 1: \textit{Fixed coefficients}.] Table \ref
{multitesthypbetaresults-fixed} summarizes an experiment involving 12
%
%t1 #&#
%
\begin{table}
\tabcolsep=0pt
\caption{Experiment 1, fixed
coefficients. SSE: sum of squared errors in the estimate of the $\bbeta
$ sequence. FP: false positive declarations in the estimate of
$\bbeta
$ sequence. FDR: realized false-discovery rate. Laplace: posterior
median estimator from the empirical Bayes procedure of
Johnstone and Silverman
(\protect\citeyear{johnstonesilverman2004}). The numbers in parentheses
indicate, in
order, the choices of $a$ and $b$ the HIB model}
\label{multitesthypbetaresults-fixed}
{\fontsize{8.5pt}{11pt}\selectfont{
\begin{tabular*}{\tablewidth}{@{\extracolsep{\fill}}lc d{2.1}d{2.1}d{2.1}d{2.1}
d{3.1}d{3.1}d{3.1}d{3.1} d{3.1}d{3.1}d{3.1}d{3.1}@{}}
\hline
& & \multicolumn{12}{c@{}}{\textbf{Number nonzero out of 1,000 means}}
\\[-4pt]
& & \multicolumn{12}{c@{}}{\hrulefill} \\
& & \multicolumn{4}{c}{\textbf{5}} & \multicolumn{4}{c}{\textbf{50}}
& \multicolumn{4}{c@{}}{\textbf{100}} \\[-4pt]
& & \multicolumn{4}{c}{\hrulefill} & \multicolumn{4}{c}{\hrulefill}
& \multicolumn{4}{c@{}}{\hrulefill} \\
\multicolumn{2}{r}{\hspace*{4pt}\textbf{Value:}} & \multicolumn
{1}{c}{\textbf{3}}
& \multicolumn{1}{c}{\textbf{4}} & \multicolumn{1}{c}{\textbf{5}}
& \multicolumn{1}{c}{\textbf{7}} & \multicolumn{1}{c}{\textbf{3}}
& \multicolumn{1}{c}{\textbf{4}} & \multicolumn{1}{c}{\textbf{5}}
& \multicolumn{1}{c}{\textbf{7}} & \multicolumn{1}{c}{\textbf{3}}
& \multicolumn{1}{c}{\textbf{4}} & \multicolumn{1}{c}{\textbf{5}}
& \multicolumn{1}{c@{}}{\textbf{7}} \\
\hline
SSE & Laplace & 35.1 & 32.8 & 17.9 & 8.5 & 210.5 & 150.8 & 99.7 & 71.9
& 331.1 & 248.3 & 177.5 & 142.9 \\
%& $(2,2)$ & 32.7 & 31.3 & 22.1 & 15.2 & 198.8 & 180.0 & 154.2 & 123.5
%& 334.1 & 321.5 & 284.8 & 240.1 \\
%&$(2,1)$ & 32.7 & 32.5 & 24.2 & 16.7 & 207.1 & 196.8 & 170.3 & 128.6 &
%356.7 & 366.3 & 319.6 & 250.3 \\
%&$(2,0.5)$ & 33.1 & 34.3 & 26.5 & 18.2 & 224.6 & 232.4 & 215.1 & 142.1
%& 354.3 & 350.6 & 310.3 & 265.1 \\
&$(1,2)$ & 35.4 & 31.9 & 17.9 & 10.3 & 205.4 & 157.7 & 116.7 & 90.6 &
334.6 & 268.2 & 213.2 & 180.4 \\
&$(1,1)$ & 35.0 & 31.3 & 18.5 & 11.1 & 200.5 & 161.9 & 124.7 & 95.3 &
329.1 & 280.8 & 229.3 & 188.5 \\
&$(1,0.5)$ & 34.7 & 31.0 & 19.6 & 12.2 & 199.6 & 170.7 & 135.3 & 100.6
& 335.2 & 302.2 & 248.1 & 196.3 \\
&$(0.5,2)$ & 37.9 & 36.8 & 18.3 & 7.3 & 242.6 & 167.3 & 104.0 & 70.8 &
395.3 & 272.8 & 182.8 & 145.7 \\
&$(0.5,1)$ & 37.6 & 36.3 & 18.1 & 7.6 & 234.9 & 164.1 & 105.0 & 72.6 &
379.5 & 268.8 & 186.4 & 148.9 \\
&$(0.5,0.5)$ & 37.4 & 35.7 & 17.9 & 7.9 & 227.5 & 161.1 & 106.2 & 74.2
& 363.6 & 266.2 & 190.9 & 151.9\\
[4pt]
FP & Laplace & 0.8 & 1.0 & 0.8 & 0.4 & 16.1 & 11.3 & 7.6 & 4.2 & 53.3 &
28.7 & 17 & 8.9 \\
%& $(2,2)$ & 0.5 & 1 & 1.3 & 1.2 & 16 & 25 & 23 & 17.1 & 83 & 129.6 &
%107.1 & 61.9 \\
%&$(2,1)$ & 0.6 & 1.5 & 1.7 & 1.6 & 82 & 121.3 & 90.4 & 38.5 & 900 &
%900 & 900 & 622.1 \\
%&$(2,0.5)$ & 1.4 & 3.1 & 3.4 & 2.6 & 941 & 950 & 950 & 706 & 900 & 900
%& 900 & 900 \\
&$(1,2)$ & 0.2 & 0.3 & 0.6 & 0.5 & 4.0 & 6.9 & 6.6 & 5.5 & 12.2 & 18.2
& 17.2 & 13.2 \\
&$(1,1)$ & 0.2 & 0.4 & 0.7 & 0.5 & 6.4 & 10.2 & 9.4 & 6.9 & 23.7 & 34.0
& 29.2 & 18.7 \\
&$(1,0.5)$ & 0.3 & 0.6 & 0.8 & 0.7 & 13.5 & 21.1 & 16.9 & 9.8 & 153.5 &
199.8 & 90.0 & 33.4 \\
&$(0.5,2)$ & 0.1 & 0.1 & 0.1 & 0.2 & 1.1 & 2.5 & 2.2 & 2.2 & 2.9 & 5.5
& 5.4 & 5.1 \\
&$(0.5,1)$ & 0.1 & 0.1 & 0.2 & 0.2 & 1.4 & 3.0 & 2.7 & 2.5 & 3.7 & 7.1
& 6.7 & 5.9 \\
&$(0.5,0.5)$ & 0.1 & 0.1 & 0.2 & 0.2 & 1.7 & 3.7 & 3.1 & 2.8 & 5.5 &
9.5 & 8.6 & 6.8 \\
[4pt]
FDR & Laplace & 0.2 & 0.2 & 0.1 & 0.1 & 0.3 & 0.2 & 0.1 & 0.1 & 0.4 &
0.2 & 0.1 & 0.1 \\
%0.2 & 0.2 & 0.2 & 0.2 & 0.3 & 0.3 & 0.3 & 0.2 & 0.5 & 0.6 & 0.5 & 0.4
%& \\
%0.2 & 0.2 & 0.2 & 0.2 & 0.6 & 0.7 & 0.6 & 0.4 & 0.9 & 0.9 & 0.9 & 0.8
%& \\
%0.3 & 0.3 & 0.3 & 0.3 & 0.9 & 1 & 1 & 0.9 & 0.9 & 0.9 & 0.9 & 0.9 & \\
&$(1,2)$ & 0.1 & 0.1 & 0.1 & 0.1 & 0.1 & 0.1 & 0.1 & 0.1 & 0.1 & 0.2 &
0.1 & 0.1 \\
&$(1,1)$ & 0.1 & 0.1 & 0.1 & 0.1 & 0.2 & 0.2 & 0.2 & 0.1 & 0.2 & 0.3 &
0.2 & 0.2 \\
&$(1,.5)$ & 0.2 & 0.1 & 0.1 & 0.1 & 0.3 & 0.3 & 0.2 & 0.2 & 0.6 & 0.6 &
0.5 & 0.2 \\
&$(0.5,2)$ & 0.1 & 0.0 & 0.0 & 0.0 & 0.0 & 0.1 & 0.0 & 0.0 & 0.1 & 0.1
& 0.1 & 0.0 \\
&$(0.5,1)$ & 0.1 & 0.0 & 0.0 & 0.0 & 0.1 & 0.1 & 0.1 & 0.0 & 0.1 & 0.1
& 0.1 & 0.1 \\
&$(0.5,0.5)$ & 0.1 & 0.0 & 0.0 & 0.0 & 0.1 & 0.1 & 0.1 & 0.1 & 0.1 &
0.1 & 0.1 & 0.1 \\
\hline
\end{tabular*}
}}
\end{table}
configurations of different sparsity patterns (5, 50, and 100 nonzero
means) and different scales (all nonzero means equal to 3, 4, 5, or 7).
\item[\textit{Experiment} 2: \textit{Random $t_3$-distributed
coefficients}.] Table \ref{multitesthypbetaresults-random} summarizes
an experiment in which the nonzero means were randomly drawn from a
heavy-tailed $t$ distribution with 5 degrees of freedom and scale
parameter $c$. We investigated 12 configurations of different sparsity
patterns (20, 50, 200, and 500 nonzero means) and different scales ($c
= 0.5, 1, 2$).
%sparsity level.] For the simulation study reported in Table
%$$
%C j^{-\alpha} , & 1 \leq j \leq50 \\
%0 , & j > 50 \\
% .
%$$
%We investigated 12 configurations, with $C\in\{10,25,50,75\}$ and $
%total magnitude and rate of decay of coefficient size.
\end{longlist}

Tables \ref{multitesthypbetaresults-fixed} and \ref
{multitesthypbetaresults-random} show the average sum of squared errors
in estimating~$\bbeta$ over 100 independent data sets. Also shown are
the average number of false positives declared by the two procedures in
each case, and the average false-discovery rate. For the
Johnstone/Silverman procedure, a false positive occurs when the
posterior median of $\beta_i$ is nonzero, but the actual value is
zero. For the Bayesian procedure using the
hypergeometric inverted-beta prior, a~false positive occurs when the
posterior inclusion probability for $\beta_i$ is greater than $50\%$
and $\beta_i$ is actually zero. This threshold reflects a 0--1 loss
function that penalizes false positives and false negatives equally,
regardless of size. A full decision-theoretic analysis incorporating
more realistic loss functions would yield a different, data-adaptive
threshold, but would only complicate the analysis slightly.

For the hypergeometric inverted-beta prior, we set $s = 0$, while $w$
and $\tau$ were estimated by importance sampling. For priors, we
assumed that $\tau\sim\C^{+}(0,\sigma)$, and that $w \sim\U(0,1)$.

In experiment 1, we used a range of values for $a$ and $b$. The best
overall choice seemed to be $a=1/2$, $b=1$, and so we focused solely on
this choice in experiment~2. Indeed, although certain alternative
choices produced improvements in specific situations, we found $a=1/2,
b=1$ to be a good all-purpose option because of its blend of good
performance in estimation and testing.

Overall, when squared error in estimation is used to decide between
procedures, our preferred Bayes procedure with $a=1/2,b=1$ wins
slightly on experiment~2, while the empirical-Bayes thresholding
procedure wins slightly on experiment 1. We attribute these differences
to the relative tail weight of the two priors. The double-exponential
prior has tails that are heavier than the Gaussian likelihood, but not
as heavy as those of the hypergeometric inverted-beta priors we
studied. This difference in tail weight becomes much more significant
in the experiment with random coefficients, since draws from a $t_3$
density produce some very large signals---much larger than signals of
size 7 in the ``fixed coefficients'' study. In experiment 2, however,
the heavier-tailed priors are wasting some of their mass in areas of
the parameter space far from the origin. Since these areas are
predestined to be unimportant by the particular choices of fixed
signals, it is no surprise that a~lighter-tailed prior such as the
double-exponential will yield superior results. Similarly, when the
coefficients are slightly larger, as in the $t_3$ signals from
experiment~2, the heavier-tailed prior will outperform.

%
%t2 #&#
%
\begin{table}
\tabcolsep=0pt
\caption{Experiment 2, random
coefficients. The HIB prior set $a=1/2$, $b=1$}
\label{multitesthypbetaresults-random}
{\fontsize{8.5pt}{11pt}\selectfont{
\begin{tabular*}{\tablewidth}{@{\extracolsep{\fill}}lc cd{2.1}d{2.1}
d{2.1}d{2.1}d{3.1} d{2.1}d{3.1}d{3.1} d{3.1}d{3.1}d{3.1}@{}}
\hline
& & \multicolumn{12}{c@{}}{\textbf{Number nonzero}}\\[-4pt]
& & \multicolumn{12}{c@{}}{\hrulefill}\\
& & \multicolumn{3}{c}{\textbf{50}} & \multicolumn{3}{c}{\textbf{100}}
& \multicolumn3{c}{\textbf{200}} & \multicolumn{3}{c@{}}{\textbf{500}}
\\[-4pt]
& & \multicolumn{3}{c}{\hrulefill} & \multicolumn{3}{c}{\hrulefill} &
\multicolumn{3}{c}{\hrulefill} & \multicolumn{3}{c@{}}{\hrulefill}\\
\multicolumn{2}{r}{\hspace*{5pt}\textbf{Scale $\bolds{c}$:}}
& \multicolumn{1}{c}{\textbf{0.5}} & \multicolumn{1}{c}{\textbf{1}}
& \multicolumn{1}{c}{\textbf{2}} & \multicolumn{1}{c}{\textbf{0.5}}
& \multicolumn{1}{c}{\textbf{1}} & \multicolumn{1}{c}{\textbf{2}}
& \multicolumn{1}{c}{\textbf{0.5}} & \multicolumn{1}{c}{\textbf{1}}
& \multicolumn{1}{c}{\textbf{2}} & \multicolumn{1}{c}{\textbf{0.5}}
& \multicolumn{1}{c}{\textbf{1}} & \multicolumn{1}{c@{}}{\textbf{2}} \\
\hline
SSE & HIB & 8.3 & 16.0 & 55.4 & 28.8 & 53.2 & 125 & 90.2 & 235 & 336 &
181 & 391 & 604 \\
& Laplace & 8.6 & 16.1 & 60.4 & 29.5 & 57.3 & 136 & 93.1 & 250 & 370 &
180 & 394 & 646 \\
[4pt]
FP & HIB & 0.0 & 0.0 & 0.2 & 0.0 & 0.2 & 0.5 & 0.1 & 0.8 & 3.3 & 0.1 &
0.9 & 10.8 \\
& Laplace & 0.4 & 3.7 & 1.1 & 2.3 & 1.6 & 3.2 & 23.5 & 34 & 19.1 & 138
& 134 & 71.5\\
\hline
\end{tabular*}
}}
\end{table}

But when the measuring stick is the false-positive rate, the fully
Bayes procedure with smaller values of $a$ and $b$ wins. It produces
far fewer false positives across the board, along with lower
false-discovery rates (suggesting that it is not merely more
conservative across the board in declaring an observation to be a
signal). It therefore seems like the more robust choice. For situations
when estimation is the goal, its performance is roughly comparable to
the existing Johnstone/Silverman procedure. Yet for situations when
testing is the goal, the Bayes procedure appears more trustworthy.

%s4 #&#
\section{Testing for superior historical performance}
\label{sec4}

%s4.1 #&#
\subsection{Data preprocessing}

Before applying our multiple-testing method, we preprocessed the data
as follows. Let $y_{it}$ be the raw data point for company $i$ in year
$t$. We first standardized the data to have zero mean and unit variance
across all countries and years. Using Bayesian treed-regression
software [\citet{gramacylee2008}], we then estimated a conditional mean~$m_{it}$
and a conditional standard deviation $s_{it}$, representing
the expected distribution of performance for other firms in company
$i$'s peer group in year~$t$. As covariates, we used a company's
industry, size, leverage, country of operation, and market share. For
an extensive discussion of how this issue relates to the disambiguation
of so-called ``Schumpeterian'' rents from ``monopolistic'' rents, see
\citet{hendersonetal2009}.

The regression-tree approach allows us to account for the highly
nonlinear, conditionally heteroskedastic relationships present in the
data. An instructive comparison can be found in Figure \ref
{examples1}, which shows three firms: JPMorgan Chase, IBM, and Gap
Instrument Corporation. It is clear that the three firms have
noticeably different peer-group means, and drastically different
peer-group standard deviations. The left-hand plots show the actual
performance, along with the ``benchmark distribution''---that is, the
mean and standard deviation of that year's expected performance, given
firm-level covariates. The right-hand plots show the performance with
respect to the benchmark distribution, all on a common normal-CDF
scale. Supplemental files available upon request from the authors show
the results of an extensive exploratory analysis of ROA versus
important covariates, and substantiates our claim that nonlinear,
conditionally heteroskedastic regression is essential here.

We then computed a $z$-score $z_{it} = (y_{it} - m_{it})/s_{it}$ for
each company-year data point. We emphasize that the term $m_{it}$
accounts only for the effects of covariates, and does not include a
random effect specific to the firm in question. Therefore, if firm $i$
systematically performs $\mu_i$ standard deviations above (or below)
its peer-group mean, and each year's performance is conditionally
independent given $\mu_i$, then
\[
(z_{it} \mid\mu_i) \sim\N(\mu_i, 1) \qquad\mbox{for $i = 1,
\ldots, n_i$.}
\]
If $\mu_i = 0$, then the sample mean of the $z_{it}$'s for firm $i$ is
normally distributed with mean $0$ and variance $1/n_i$, where $n_i$ is
the number of observations we have for that firm (ranging from 5 to
43). This is our preliminary null hypothesis. Stated in an equivalent form,
\[
z_i = \bar{z}_i \sqrt{n_i} \sim\N(0,1) .
\]
These $z$-scores are the raw inputs to our multiple-testing approach.
Based on the simulation results above, we are reporting results for
$a=1/2, b=1$, which seemed to provide the best overall results in terms
of testing.

%s4.2 #&#
\subsection{Summary of results}

We ran the proposed multiple-testing method on the cohort of firms for
which at least 5 years of past data were available. This initial sieve
left us with a cohort of 37,014 firms, each with somewhere between 5
and 43 annual observations.

Of the tested cohort, 1,076 firms (or about $3\%$) had posterior
probabilities of outperformance larger than $90\%$, indicating moderate
to high confidence that they have systematically outperformed their
peer groups. For this cohort, the expected group-wise false discovery
rate (FDR) is $2\%$; this can be computed by simply averaging the
posterior probabilities that each firm in the cohort comes from the
null model. An additional 705 firms had posterior probabilities of
outperformance between $50\%$ and $90\%$. For this intermediate group,
the expected FDR is $28\%$.

The top 10 overall firms ranked by posterior probability are described
in Table~\ref{tabtop10}, along with the reason that firm dropped out
of the database (if applicable). Of these 10 firms, 8 seemed to
outperform their peer group, while 2 seemed to underperform. The first
non-American firm on the list is British--American Tobacco,
incorporated in (of all places) Malaysia, which ranks 11th by estimated
posterior inclusion probability.

%
%t3 #&#
%
\begin{table}
\tabcolsep=3pt
\caption{Ten firms with the highest posterior
probabilities of having a nonzero mean}
\label{tabtop10}
\begin{tabular*}{\tablewidth}{@{\extracolsep{\fill}}llc@{}}
\hline
\textbf{Company} & \multicolumn{1}{c}{\textbf{Description}}
& \multicolumn{1}{c@{}}{\textbf{Books}} \\
\hline
Alfacell Corporation & A biotech firm specializing in RNA-based
technologies. & --- \\
Wyeth & Large drug company; recently bought out by Pfizer. & --- \\
American List Corp & Bulk mailing firm. Bought out in 1997. & --- \\
Deluxe Corp & Financial and logistical services for small businesses. &
--- \\
Tambrands & Personal hygiene products. Bought out in 1997. & --- \\
Toth Aluminum & Developed aluminum technology. Defunct. & --- \\
UST & A tobacco holding company. Bought out in 2009. & --- \\
WD-40 & Manufactures the anticorrosive and lubricating agent. & --- \\
Landauer & Specializes in services relating to radiation safety. & ---
\\
Merck & Large drug company. & BTL, ISE\\
\hline
\end{tabular*}
\end{table}

\begin{figure}[t!]

\includegraphics{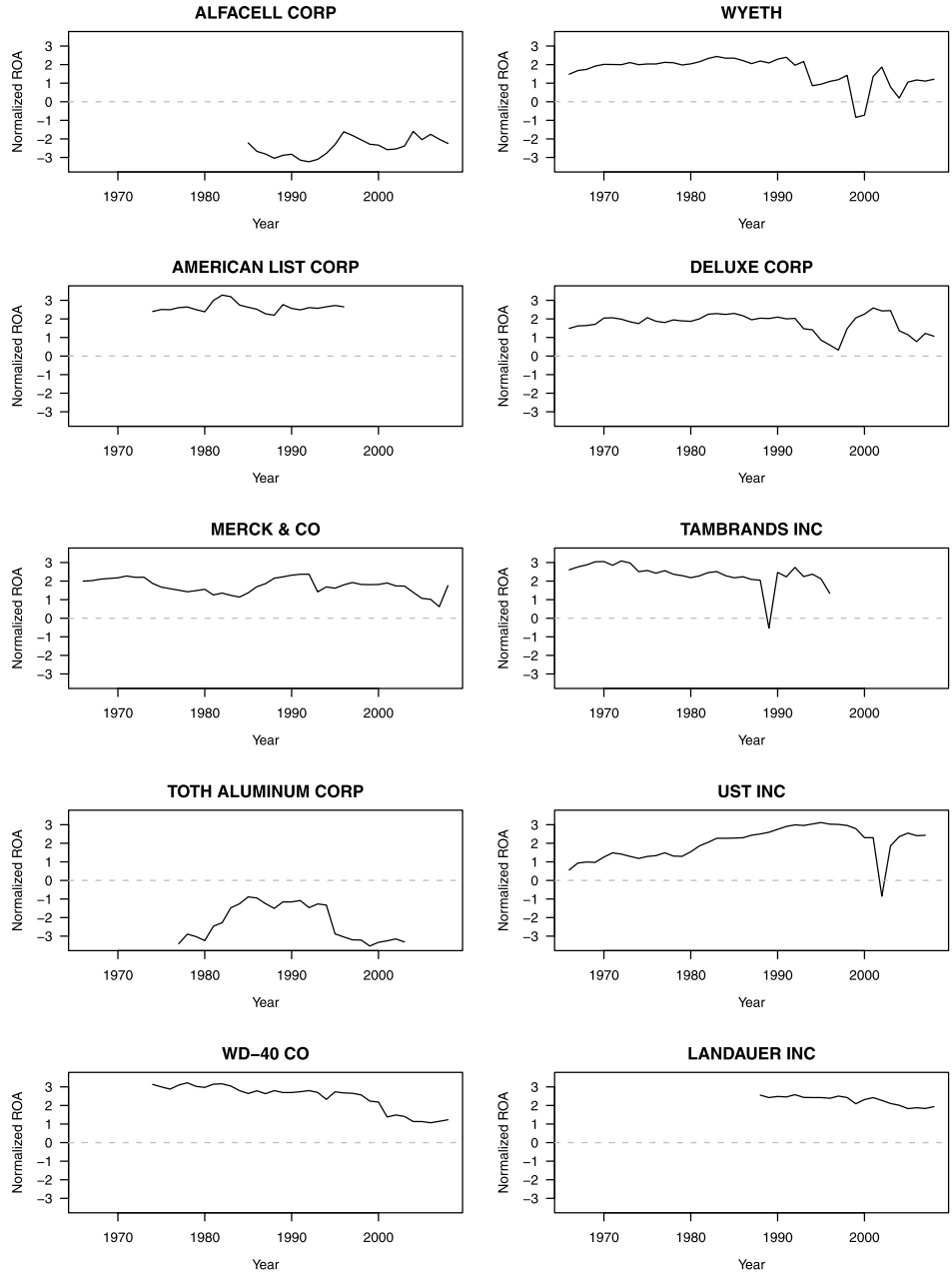}

\caption{The performance trajectories for the ten firms with
the highest posterior probabilities of having a nonzero mean.}
\label{figtop10initial}
\vspace*{6pt}
\end{figure}

The historical trajectories for these 10 firms can be seen in Figure
\ref{figtop10initial}. Two are large drug companies; the rest come
from a variety of different industries. All but four---Wyeth, Merck,
Tambrands, and WD-40---are likely unknown to the average consumer.

%
%f2 #&#
%

These results are best thought of as a reasonable upper bound to the
actual number of sustained superior performers. This is true for at
least two reasons. First, although we used all data for 53,038 firms to
fit the regression tree models and compute $m_{it}$ and $s_{it}$, we
did not conduct hypothesis tests for the 16,024 firms with less than 5
years of data. It is difficult to know what ``long-term superiority''
even means for this vast group of firms with so short a history.
Moreover, their presence in the testing stage of the analysis would
likely bias the estimate of $w$ (the prior inclusion probability)
downward, because the Bayes factor so strongly favors the null
hypothesis for such a short trajectory. (This results from the
well-known Bayesian ``Occam's razor'' effect that arises when comparing
models of different dimensionality.) This introduces a possible
survivorship bias into our procedure. But given the assumption of
exchangeability in our model, we believe that the effects of
survivorship bias are less severe than the likely effects of watering
down the cohort with so many firms for which the null hypothesis is so
likely a~priori.

Second, and more importantly, our analysis assumes that a company's ROA
result in year $t$ is independent of results from previous years, given
the peer group mean and standard deviation. This is unlikely to be
exactly true, and therefore introduces an upward bias in our estimate
of the number of superior performers (due to the fact that
autocorrelation reduces the effective sample size available for testing $H_0$).

One way of accounting for this bias is to introduce specific parametric
assumptions about the nature of a ``true null'' trajectory. Indeed,
this is an active and promising area of research in both this and in
parallel fields (e.g., time-course microarray data). Our focus on this
paper, however, is on large-scale screening with relatively few
assumptions. We therefore eschew explicit parametric longitudinal
models and adopt the following alternative strategy in an attempt to
get a fast, crude assessment of how the independence assumption may
affect our results:
\begin{longlist}[(1)]
\item[(1)] For each firm in the testing cohort, we estimate a one-lag
autocorrelation coefficient, $\hat{\phi}_i$. For the handful of firms
for which this estimate is negative, we threshold at zero, since we do
not wish to introduce negative correlation into the sampling
distribution for the data.
\item[(2)] We compute an effective sample size for each trajectory as
\[
\hat{n}_i = n_i \cdot\biggl( \frac{1-\hat{\phi}_i}{1 + \hat{\phi
}_i} \biggr)
\]
using the well-known correction for autocorrelation. While this is
motivated by simple AR(1)-type null models, one may interpret the
multiplicative term involving $\hat{\phi}$ purely as a deflator,
corresponding to the reduction in information in each longitudinal
sample compared to the i.i.d. case.
\item[(3)] We recompute the $z$-score as $\hat{z}_i = \bar{z}_i \sqrt
{\hat{n}_i}$.
\end{longlist}
We then repeat the testing procedure using the $\hat{z}_i$'s as data,
which has the effect of inflating the variance under the null
hypothesis. This correction led to 262 firms with a posterior
probability greater than $90\%$ (expected FDR for the group: $2\%$),
and an additional 222 with a posterior probability between $50\%$ and
$90\%$ (expected FDR for the group: $26\%$). The top 10 firms remained
unchanged, except for Toth Aluminum and Alfacell.

Our results appear to be qualitatively similar to those of
\citet{hendersonetal2009}, who use essentially the same data. But
we will
point to two important methodological differences that likely account
for any major divergence in testing outcomes. First, we use
model-averaged estimates from Bayesian treed regression to estimate a
conditional mean and standard deviation for every company in every
year. In contrast, \citet{hendersonetal2009} use linear quantile
regression, which is a~fundamentally different---and arguably less
flexible---way of accounting for conditional heteroskedasticity (which
appears to be the dominant effect of covariates). Second, we adjust
each company's longitudinal results individually to account for
firm-level heterogeneity with respect to autocorrelation. In contrast,
\citet{hendersonetal2009} account for longitudinal dependence by
assuming that the same semi-parametric Markov model holds across the
entire population of ``lucky'' firms.

%s4.3 #&#
\subsection{Comparison with the popular literature on corporate success}

As a~small aside, it is interesting to compare these results to the
conclusions of a~handful of well-known books that purport to explain
corporate success. We took a small, nonscientific sample of these
books, in an attempt to gauge whether the results from the
multiple-testing model correspond to widely held notions about
successful firms. Table \ref{tabbookchoices} briefly describes these
books, and indicates whether the basis for selecting the study cohort
was qualitative or quantitative in nature. The books were chosen in
conjunction with a group of senior management consultants at Deloitte
Consulting, who judged the list to be fairly representative of the
popular literature.

%
%t4 #&#
%
\begin{table}
\tabcolsep=3pt
\caption{The popular books selected for comparison}
\label{tabbookchoices}
\begin{tabular*}{\tablewidth}{@{\extracolsep{\fill}}lc p{154pt} c@{}}
\hline
\textbf{Title} & \multicolumn{1}{c}{\textbf{Published}} &
\multicolumn{1}{c}{\textbf{Selection method}}
& \multicolumn{1}{c@{}}{\textbf{Basis}} \\
\hline
Good to Great &2001 & Companies from 1965--1981 selected on the basis
of shareholder return & Quantitative \\
Built to Last & 1994 & Companies founded before 1950 that met certain
success criteria & Qualitative \\
In Search of Excellence & 1982 & Surveys of executives at
author-selected firms & Qualitative \\
Competitive Strategy & 1980 & Author selected examples to support
theory; method unclear & Qualitative \\
Hidden Values & 2000 & Author selected examples to support theory;
method unclear & Qualitative \\
Blueprint to a Billion & 2006 & Time to achieve $\$1$ billion in
revenue after initial public offering & Quantitative \\
What Really Works & 2003 & Correspondence with prespecified ``top
management practices'' & Qualitative \\
Stall Points & 2008 & Patterns of stalls and recovery in revenue growth
& Quantitative \\
Blue Ocean Strategy & 2005 & Author selected examples to support
theory; method unclear & Qualitative\\
\hline
\end{tabular*}
\end{table}

These books follow a common recipe: start with a group of companies;
identify the ``successful'' ones; look for patterns in their behavior;
and abstract those behaviors into a small set of principles that can
tell others how to run their businesses better. One important
difference between these books and the approach considered here is the
choice of outcome variable. In some books the outcome variable is
multidimensional, and therefore richer than our choice of ROA. Thus,
while comparisons are instructive, they do not support the conclusion
that our study is objectively right and the others wrong. Moreover, as
a referee observed, the authors of these books may have different
things in mind when they define success.

Yet, collectively, these studies exhibit many unacknowledged sources of
bias, which our study attempts to address. None, for example, make a
serious attempt to verify statistically that the selected companies
have done anything special when compared with a suitable reference
population. This opens up the possibility that they have been studying
companies that were lucky, rather than great---the precise null
hypothesis considered in this paper. There are also serious issues with
selection bias---both in terms of metric selection and of company
selection---and of survivorship bias (although our study is also
imperfect in this regard).

Perhaps for these reasons, serious discrepancies emerged between the
popular literature and the conclusions of the multiple-testing
procedure considered here. Across the nine books considered, there were
209 distinct firms that were used as case studies---some positive, some
negative---and that also appeared in our cohort of firms with 5 or more
years of data. Of the top ten firms flagged in the previous section,
only one was mentioned in any of the 9 books: Merck, a case study in
Built to Last (BTL) and In Search of Excellence (ISE). Of the 209 firms
collectively mentioned in these books, only 9 appear on our list of
firms with ROA trajectories significantly better than those of their
peer groups, once longitudinal dependence is accounted for.

%s5 #&#
\section{Final remarks}

We have developed a Bayesian multiple-testing procedure based upon a
heavy-tailed prior for the nonzero means. These priors form an
interesting, novel class of normal variance mixtures, the
hypergeometric inverted-beta class. Overall, the procedure has the
nice theoretical property of a redescending score function under the
alternative model, and seems to perform as well as, or better than,
existing gold-standard methods. Moreover, it allows relevant Bayesian
and frequentist summaries to be computed with minimal computational
fuss. This property arises from the simple, known form of the marginal
distribution $m(y)$.

We have applied the method to a large data set on historical corporate
performance, and compared the results of our analysis to some popular
books that deal with the same subject. These books appear to be
studying a sample where the large majority of firms have ROA
performance profiles that are statistically indistinguishable from
luck. Meanwhile, there on the order of hundreds of firms (out of a
group of over 37,000) whose performance is at least suggestive of a
sustained advantage, and yet were not considered in these high-profile
case studies.

%
%apA #&#
%
\begin{appendix}\label{app}
%sB #&#
\section{The proposed family of priors}\label{appA}

%sB.1 #&#
\subsection{Connection with classical shrinkage rules}

Our new class of priors has its genesis in the large body of work on
classical shrinkage rules, where a~multivariate normal prior $\bbeta
\sim\N(0, \lambda^2 I)$ is assumed, where $\bbeta= (\beta_1,
\ldots, \beta_p)$. Many common estimators for this problem, both
Bayesian and non-Bayesian, are of the form $\hat{\bbeta} (\by) = \{1
- g ( Z ) \} \by$ for $Z = \Vert\by\Vert^2$ [e.g.,
\citet{jamesstein1961}, \citet{strawderman1971},
\citet{stein1981}, \citet{fourdrinieretal1998}]. The
central issue is how to identify ``nice'' functions $g(Z)$, and how to
understand priors for global variance components in terms of the
behavior of the estimators they yield.

The constraint to rationality---that is, the requirement that there
exists a~prior $\p(\kappa)$ such that, for all $Z$, $g(Z) = E( \kappa
\mid Z ) $ under the posterior $p( \kappa\mid Z ) $---rules out a wide
class of potential estimators. The function $g(Z)$ cannot, for example,
be a polynomial of order two or greater. Indeed, the functional form of
a $g(Z)$ that respects admissibility will typically be quite complicated.

It is natural to look in the class of estimators where
$g(Z)=p(Z)/q(Z)$, a~ratio of power-series expansions. One can construct
such a $g(Z)$ by assuming that $(\bbeta\mid\lambda^2) \sim\N(0,
\lambda^2 I)$, and then defining $ \hat{\bbeta} (\lambda^2) = \E(
\bbeta\mid\lambda^2 , \by) $. After removing the dependence upon
$\lambda^2$ by marginalizing, this leads to
\[
\hat{\bbeta} = E_{\lambda^2 \mid\by} \{ \hat{\bbeta} ( \lambda
^2 ) \} = \{ 1 - E( \kappa\mid Z ) \} \by,
\]
recalling that $\kappa= 1/(1 + \lambda^2)$. We can therefore identify
$g(Z)$ with $E( \kappa\mid Z ) $, the posterior expectation of $\kappa
$, given $Z$.

One can define a class of priors for $\kappa$ indexed by $(a,b,\tau
,s) $, which we call the hypergeometric inverted-beta class, such that
%
%eB.1 #&#
%
\begin{eqnarray}\quad
\label{HBglobalestimatorequation}
g(Z) &=& E( \kappa\mid Z ) \nonumber\\[-8pt]\\[-8pt]
&=& \frac{a + p/2}{a + b + p/2} \frac{\Phi_1(b,
1; a + b + p/2 + 1; s + Z /2, 1-1/\tau^2)} { \Phi_1(b, 1; a + b +
p/2; s + Z /2, 1-1/\tau^2)} ,\nonumber
\end{eqnarray}
where $a$, $b$, and $\tau$ are positive real numbers; $s$ is any real
number; and $\Phi_1$ is the degenerate hypergeometric function of two
variables [\citet{gradshteynryzhik1965}, Equations
9.261.1--9.261.3].

This $g$ is a ratio of power series, and can be computed quite rapidly
for a given tuple $(a,b,\tau,s)$ and a given $Z$. It leads to a large
class of admissible estimators with a wide range of possible behavior.
In particular, it includes many estimators that exhibit robustness to
large values of Z; many estimators that offer significant risk
reduction near $Z = 0$; and many that do both. This class generalizes
the form noted by \citet{maruyama1999}, which contains the
positive-part James--Stein estimator as a limiting (improper) case.

%sB.2 #&#
\subsection{Hypergeometric inverted-beta priors}

The connection with multiple testing is as follows. Recall that
under
the alternative model, $\beta_i$ is conditionally\vadjust{\goodbreak} normal with variance
$\lambda_i^2$. Our approach is to work with the transformed variable
$\kappa_i = 1/(1 + \lambda_i^2)$, and to define the following prior
for~$\kappa_i$. Suppressing subscripts for the moment,
%
%eB.2 #&#
%
\begin{equation}
\label{generalhypbetaequation}
\p(\kappa) = C^{-1} \kappa^{a - 1} (1 - \kappa)^{b - 1}
\biggl\{ \frac{1}{\tau^2} + \biggl(1 - \frac{1}{\tau^2} \biggr) \kappa
\biggr\}^{-1} \exp(-s \kappa) ,
\end{equation}
where $a, b, \tau> 0$ and $s \in\mathbb{R}$, and where $C_1$ is a
constant of proportionality. We denote the hypergeometric-beta prior on
the $\kappa$ scale by $\kappa\sim\HB(a,b,\tau,s)$.

The normalizing constant
%
%eB.3 #&#
%
\begin{equation}
\label{hypergeometricintegral}
C = \int_0^{1} \kappa^{a - 1} (1 - \kappa)^{b - 1} \biggl\{
\frac{1}{\tau^2} + \biggl(1 - \frac{1}{\tau^2} \biggr) \kappa
\biggr\}^{-1} \exp(-s \kappa) \,\dd\kappa
\end{equation}
can be computed using hypergeometric series. Using the theory laid out
in \citet{gordy1998} and \citet{PolsonScott2010c}, we get
%
%eB.4 #&#
%
\begin{equation}
\label{normalizingconstantphi1}
C = e^{-s} \Be(a, b) \Phi_1(b, 1, a + b, s, 1-1/\tau^2) ,
\end{equation}
where $\Phi_1$ is the degenerate hypergeometric function of two
variables [\citet{gradshteynryzhik1965}, 9.261]. This function can be
calculated accurately and rapidly by transforming it into a convergent
series of $\kummertwo$ functions [Section 9.2 of
\citet{gradshteynryzhik1965}, \citet{gordy1998}], making
evaluation of (\ref
{normalizingconstantphi1}) quite fast for most allowable choices of the
parameters.

The implied density for $\lambda_i^2$ takes the form
%
%eB.5 #&#
%
\begin{equation}
\label{MTIBdensity}\qquad
p(\lambda^2) = C^{-1} (\lambda^2)^{b-1} (\lambda^2 + 1)^{-(a+b)}
\exp\biggl\{ -\frac{s}{1+\lambda^2} \biggr\} \biggl\{ \tau^2 +
\frac{1-\tau^2}{1+\lambda^2} \biggr\}^{-1} .
\end{equation}
This is a generalization of the inverted-beta distribution, also known
as Pearson's type VI distribution. Indeed, it reduces to an inverted
beta in the special case where $s=0, \tau=1$, in which case $a\lambda
^2/b$ will follow an $F(2b,2a)$ density.

The hypergeometric inverted-beta family contains many well-known
subfamilies of priors for $\kappa$. These include the beta
distribution, the generalized beta distribution
[\citet{mcdonaldxu1995}], and the Gauss hypergeometric
distribution [\citet{armerobayarri1994}]. The family is itself
contained in the class of compound confluent hypergeometric
distributions [\citet{gordy1998}], which has two extra parameters that
are not relevant in this context. These various related families are
why we call (\ref{MTIBdensity}) the hypergeometric inverted-beta prior.
The transformed density on the $\kappa$ scale resembles a beta
distribution, and we call this family the hypergeometric-beta (HB)
prior.

The family in (\ref{generalhypbetaequation}) has one major advantage
over other similar priors: there exist easily computable expressions
for the posterior mean $\E(\beta_i \mid y_i)$ and the marginal
density $m_1(y_i) = \int\N(y_i \mid\beta_i, \sigma^2) \p(\beta
_i) \,\dd\beta_i$ under the hypothesis that $\beta_i \neq0$. We
derive these expressions in Appendix \ref{appB}.

%sB.3 #&#
\subsection{Shrinkage profiles}

%are given up to constant terms.}
%Prior for $\beta_i$ & Prior for $\lambda_i$ & Prior for $\kappa_i$ \\
%Double-exponential & $\lambda_i \ \exp\{\lambda_i^2 / 2 \}$ & $
%Cauchy & $\lambda_i^{-2} \ \exp(-1/2\lambda_i^2)$ & $\kappa_i^{ -
%( 1 - \kappa_i )}}$ \\
%Strawderman--Berger & $\lambda_i \ (1 + \lambda_i^2)^{-3/2}$ & $
%%Normal--Exponential--Gamma & $\p(\lambda^2) \propto(1 +
%%\\
%Horseshoe & $(1+\lambda_i^2)^{-1} $ & $\kappa_i^{-1/2} (1-

We now turn to the specification of the four hyperparameters, and to
the different ``local shrinkage profiles'' that are accessible through
different choices of these parameters.\vadjust{\goodbreak}

All normal scale-mixtures have an implied shrinkage profile $\p(\kappa
_i)$, which describes the amount of shrinkage toward the origin that is
expected a priori. The prior's behavior near $\kappa_i = 0$
controls the tail weight of the marginal prior for $\beta_i$, while
the behavior near $\kappa_i = 1$ controls the strength of shrinkage
near zero.

%
%f3 #&#
%
\begin{figure}

\includegraphics{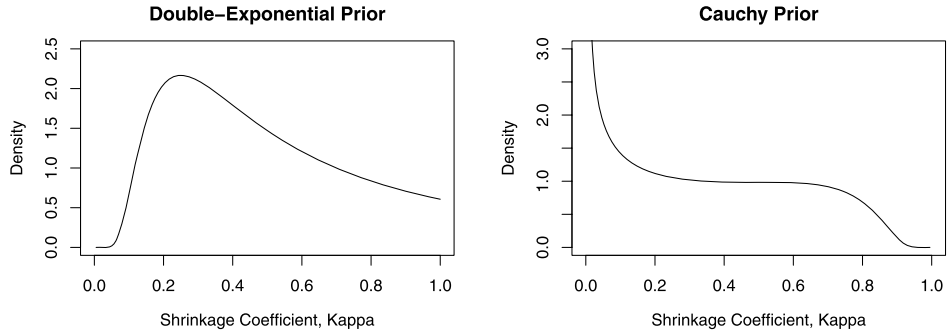}

\caption{Implied shrinkage profiles for double-exponential
and Cauchy priors.}
\label{otherkappafigure}
\end{figure}

\begin{figure}

\includegraphics{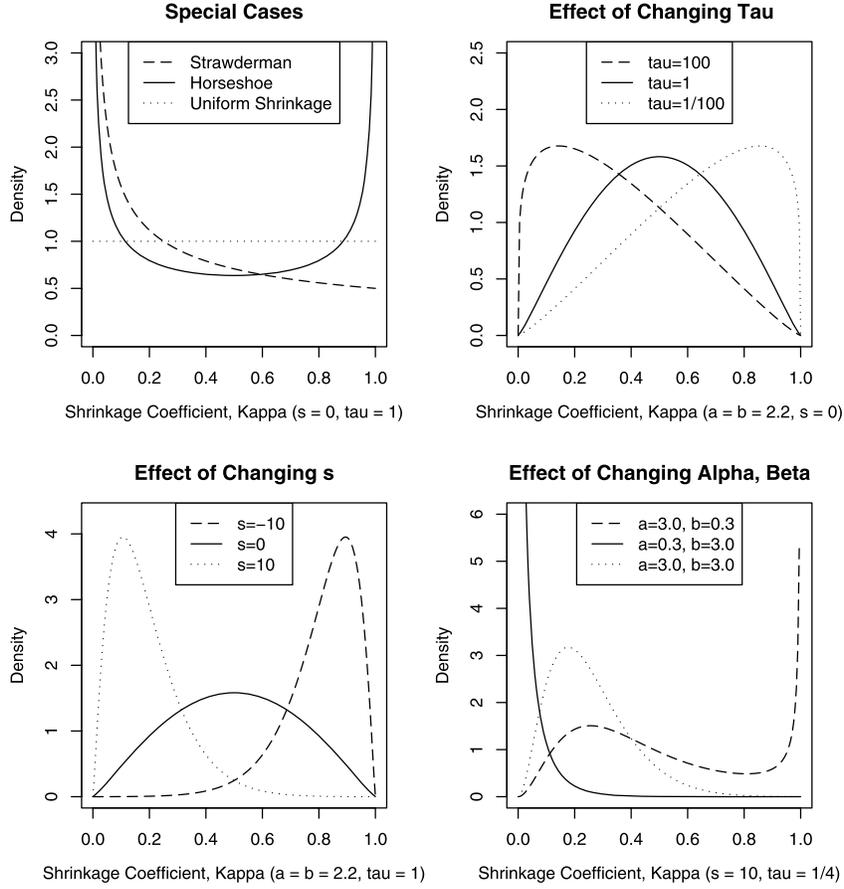}

\caption{Effect of changing the four parameters $(a, b, s,
\tau)$ on the density for the shrinkage coefficient $\kappa$.}
\label{effect-of-parchangesfigure}
\end{figure}

Figure \ref{otherkappafigure} plots the implied shrinkage profiles for
two common priors: the double-exponential and Cauchy priors. Contrast
these shrinkage profiles with the wide range of shapes that are
accessible through the hypergeometric inverted-beta density, some of
which are shown in Figure \ref{effect-of-parchangesfigure}.

%
%f4 #&#
%

One important special case of the hypergeometric inverted-beta family
is the Strawderman prior [\citet{strawderman1971}], which corresponds
to $a = 1/2$, $b=1$, $s=0$, and $\tau=1$. Another special case is the
half-Cauchy prior on the scale factor~$\lambda$, studied by
\citet{gelman2006} and \citet{CarvalhoPolsonScott2008a}. This
corresponds to
$a = b = 1/2$, $s=0$, and $\tau= 1$. Yet a third special case is the
uniform-shrinkage prior, where $a = b = 1$, $s=0$, and $\tau= 1$. All
of these can be seen in the upper-left pane of Figure \ref
{effect-of-parchangesfigure}.

Clearly, (\ref{generalhypbetaequation}) can lead to many
standard-looking shapes that are similar to other normal scale
mixtures. Yet it can also produce a wide variety of other densities
that are inaccessible through other standard families. We now describe
the role of each hyperparameter, recalling that more probability near
$\kappa= 1$ means more aggressive shrinkage.

First, $\tau$ is a global scaling factor, with larger values leading
to larger marginal variance in $\beta$. To see this, suppose that all
components of $\bbeta$ have a common variance component in addition to
their idiosyncratic ones: $(y_i \mid\beta_i) \sim\N(\beta_i,
\sigma^2)$ and $\beta_i \sim\N(0, \sigma^2 \tau^2 \lambda_i^2)$.
The form involving $\tau$ in (\ref{generalhypbetaequation}) arises
from the special case of assuming a half-Cauchy prior for each $\lambda
_i$, as in the horseshoe prior of \citet
{CarvalhoPolsonScott2008a}. The
generalization of the scaled half-Cauchy prior to arbitrary $a$, $b$,
and $s$ then arises quite naturally on the $\kappa$ scale. Shifting
$\tau$ up and down causes the shrinkage profile to be shifted left and
right, respectively, controlling the overall aggressiveness of shrinkage.

The parameters $a$ and $b$ are analogous to those of beta distribution,
to which (\ref{generalhypbetaequation}) reduces when $\tau= 1$ and $s
= 0$. Smaller values of $a$ encourage heavier tails in $\pi(\beta$),
with $a=1/2$, for example, yielding Cauchy-like tails. Smaller values
of $b$ encourage $\p(\beta)$ to have more mass near the origin, and
eventually to become unbounded; $b = 1/2$ yields, for example, $\p
(\beta) \approx\log(1+1/\beta^2)$ near $0$.

Finally, $s$ is a second global scaling factor, though with a different
effect than $\tau$ on the shape of the density. This parameter has an
interpretation as a ``prior sum of squares,'' with the caveat that it
can also be negative.

The scale parameters $\tau$ and $s$ do not control the behavior of
$\pi(\lambda)$ at $0$ and~$\infty$. Specifically, $\pi(\lambda)$
behaves like $\lambda_i^{2b-1}$ near the origin, and like $\lambda
_i^{-(2a+1)}$ in the upper tail. Since $\pi(\beta)$ has the same
polynomial rate of decay as $\pi(\lambda)$,~$a$ can be chosen to
reflect the desired tail weight of $\pi(\beta)$.

%sB.4 #&#
\subsection{The score function and overshrinkage of exceptional observations}

We recall the following theorem from
\citet{CarvalhoPolsonScott2008a}.\vspace*{-2pt}
%
%thB.1 #&#
%
\begin{theorem}
\label{postmeanlocationmodel}
Let $p(|y - \beta|)$ be the likelihood, and suppose that $\p(\beta)$
is a mean-zero scale mixture of normals: $(\beta\mid\lambda) \sim\N
(0, \lambda^2)$, with $\lambda$ having proper prior $\p(\lambda)$.
Assume further that the likelihood and $\p(\beta)$ are such that the
marginal density $m(y) < \infty$ for all $y$. Define the following
three pseudo-densities, which may be improper:
\begin{eqnarray*}
m^{\star}(y) &=& \int_{\mathbb{R}} p(|y - \beta|) \p^{\star
}(\beta) \,\dd\beta, \\[-2pt]
\p^{\star}(\beta) &=& \int_{\mathbb{R}^+} \p(\beta\mid\lambda)
\p^{\star}(\lambda) \,\dd\lambda, \\[-2pt]
\p^{\star}(\lambda) &=& \lambda^2 \p(\lambda) .
\end{eqnarray*}
Then
%
%eB.6 #&#
%
\begin{eqnarray}\label{nicksrepresentation}
E(\beta\mid y) &=& \frac{m^\star(y) }{m (y) } \,\frac{\dd}{ \dd y}
\log m^\star( y ) \nonumber\\[-9pt]\\[-9pt]
&=& \frac{1}{m (y) } \,\frac{\dd}{ \dd y} m^\star( y ) .
\nonumber\vspace*{-2pt}
\end{eqnarray}
\end{theorem}

Versions of this representation theorem appear in
\citet{masreliez1975}, \citet{polson1991} and
\citet{pericchismith1992}.
Theorem \ref{postmeanlocationmodel} relaxes a specific regularity
condition having to do with the boundedness of $\p(\beta)$, and
extends the usual result to situations where $\p(\beta)$ is a scale
mixture of normals with proper mixing density and finite marginal $m(y)$.

The theorem characterizes the behavior of an estimator in the presence
of large signals. Specifically, it says that we can achieve ``inherent
Bayesian robustness'' by choosing a prior for $\beta$ such that the
derivative of the log predictive density is bounded as a function of
$y$. Ideally, of course, this bound should converge to $0$ for large
$|y|$, and will lead to $\E(\theta\mid y) \approx y$ for large $|y|$.
This will avoid the overshrinkage of exceptional observations---clearly
an important goal in large-scale simultaneous testing problems.

It is easy to verify, using the results of the previous subsection,
that normal scale mixtures with hypergeometric inverted-beta mixing
distributions satisfy the property of tail robustness. This helps to
explain their good performance in high-dimensional settings.\vspace*{-2pt}

%sB.5 #&#
\subsection{The effect of shared shrinkage parameters}

\begin{figure}

\includegraphics{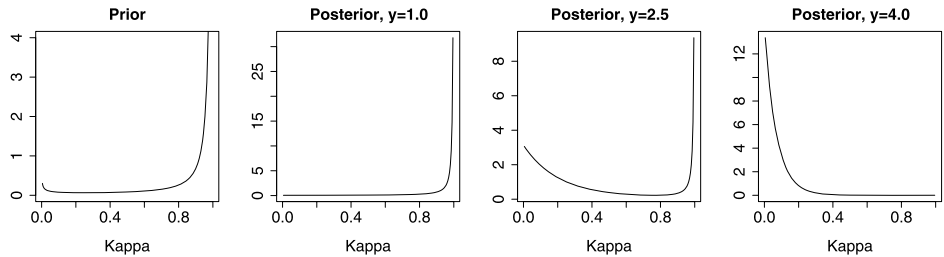}

\caption{The left pane shows the prior for $\kappa$ when
$\tau= 1/15$, $s=0$, and $a = b = 1/2$, reflecting a prior bias for
strong shrinkage. The next three panes show the different posteriors
for $\kappa$ upon observing a single data point: $y=1.0$, $y=2.5$, or
$y=4.0$, respectively.} \label{smalltau-ygrows}
\vspace*{-3pt}
\end{figure}

The hypergeometric inverted-beta prior allows a combination of global
and local shrinkage that can be both flexible and robust. Figure \ref
{smalltau-ygrows} shows how a very small value of $\tau$, encouraging
strong global shrinkage, can be reinforced by a small
observation
($y=1.0$), and yet be almost completely overruled by a large
observation ($y=4.0$).\vadjust{\goodbreak} Meanwhile, the marked bimodality for an
intermediate observation such as $y=2.5$ reflects uncertainty about
whether such an observation corresponds to signal or noise, with the
posterior mean for $\beta$ averaging over both possibilities.

%
%f5 #&#
%

This example demonstrates that global shrinkage through $\tau$ can be
very effective at squelching noise in high-dimensional problems. It is
crucial, however, that $\tau$ be estimated from the data, and that the
prior for $\kappa_i$ grow sufficiently fast near $0$ in order to allow
$\kappa_i$ to escape the strong ``gravitational pull'' of a small
$\tau$ when $y_i$ is large (as in this example when $y_i / \sigma=
4$). We recommend setting $a = 1/2$ in sparse problems involving a
normal likelihood; see \citet{CarvalhoPolsonScott2008a} for further
discussion. In situations with heavier-tailed sampling models, it may
be appropriate to choose a smaller value of $a$.

\begin{figure}

\includegraphics{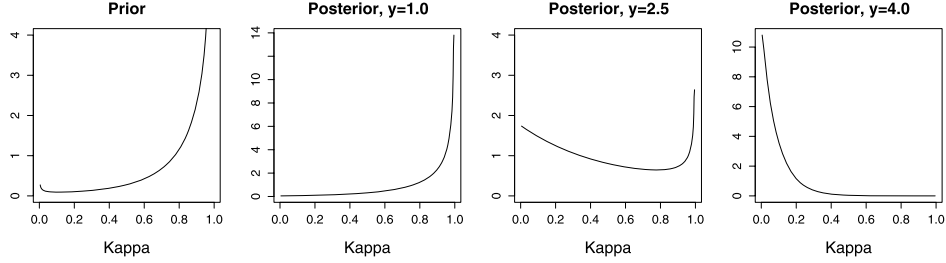}

\caption{The left pane shows the prior for $\kappa$ when
$\tau= 1$, $a = b = 1/2$, and $s=-4$. The next three panes show the
different posteriors for $\kappa$ upon observing a single data point:
$y=1.0$, $y=2.5$, or $y=4.0$, respectively.}
\label{smalls-ygrows}
\end{figure}

When $1-1/\tau^2$ is very close to $1$ (or when $1-\tau^2$ is very
close to 1 for $\tau< 1$), the $\Phi_1$ functions may become slow to
evaluate due to the slow convergence of the series representations
given in the \hyperref[app]{Appendix}. In our experience, the issue
becomes practically
significant in a serial computing environment only when $\tau^2$ is
larger than 1,000 or smaller than $1/1\mbox{,}000$. Additionally, global
shrinkage can take place through $s$ rather than $\tau$ (with $\tau$
being set equal to~$1$). Then $\kappa_i \sim\HB(a,b,\tau=1, s)$,
and so
\[
(\kappa_i \mid y_i) \sim\HB(a + 1/2, b, \tau=1, s + y_i^2/2\sigma
^2) .
\]
Figure \ref{smalls-ygrows} shows that global shrinkage through $s$ can
produce results quite similar to global shrinkage through
$\tau$.\vspace*{-2pt}

%
%f6 #&#
%

%distribution}
%
%The previous expressions show that posterior means under
%hypergeometric inverted-beta priors are easy to compute for fixed
%values of the hyperparameters. Averaging over uncertainty with respect
%to these hyperparameters, however, requires Markov-chain Monte Carlo
%to compute posterior means. This, in turn, requires a method to
%generate random draws from the hypergeometric inverted-beta
%distribution.
%
%Luckily this is straightforward using rejection sampling under a beta
%proposal. Some algebra yields the required bound on the density
%function:
%p(\kappa\mid a,b,s,\tau) &\leq& M \cdot\{ \frac{\kappa^{a-1} (1-
%M &=& \frac{ \tau^2 \cdot\max(1, e^{-s})} { e^{-s} \cdot
%The part of (\ref{HBdensitybound1}) inside braces is the density of a
%beta random variable. This suggests the following sampler:
%$$
%u \leq\frac{\min(1, \tau^2)} {\max(1, e^{-s}) } \cdot\frac{ e^{-s
%$$
%and otherwise return to Step 1.
%
%In practice, this sampler is efficient when $a$ and $b$ are both less
%than $1$, in the sense that the bound $M$ will be moderate. Better is
%to sample $\kappa\sim\Be\{\min(a,1), \min(b,1) \}$ in Step 1 of the
%algorithm, with the obvious modification of the acceptance probability
%in Step 2. This will prevent $M$ from being too large in most common
%situations.
%
%In cases where many draws with a single combination of parameters are
%required, it may be most efficient to minimize $M$ over $(a,b)$ using
%a numerical optimization routine.

%sC #&#
\section{Expressions for moments and marginals}\label{appB}\vspace*{-2pt}

Throughout this section, we suppress conditioning on $\beta_i$'s
nonzero status. Under our hypergeometric inverted-beta model, the joint
distribution for $y_i$ and $\kappa_i$ takes the form
\[
p(y_i, \kappa_i) \propto\kappa_i^{a' - 1} (1-\kappa_i)^{b-1}
\biggl\{ \frac{1}{\tau^2} + \biggl(1 - \frac{1}{\tau^2} \biggr)
\kappa_i \biggr\}^{-1}
e^{-\kappa_i s'} ,
\]
where now $s' = s + y_i^2/(2\sigma^2)$ and $a' = a + 1/2$.\vadjust{\goodbreak}

The moment-generating function of (\ref{generalhypbetaequation}) is
easily shown to be
\[
M(t) = e^t \frac{\Phi_1(b, 1, a + b, s-t, 1-1/\tau^2)}{ \Phi_1(b,
1, a + b, s, 1-1/\tau^2)} .
\]
See, for example, \citet{gordy1998}. Expanding $\Phi_1$ as a sum of
$\kummer$ functions and using the differentiation rules given in
Chapter 15 of \citet{AbraSteg1964} yields
%
%eC.1 #&#
%
\begin{equation}
\label{equationmoments}
\E(\kappa^n \mid\by, \sigma^2) = \frac{(a')_n}{(a' + b)_n} \frac
{\Phi_1(b, 1, a' + b + n, s', 1-1/\tau^2)}{ \Phi_1(b, 1, a' + b, s',
1-1/\tau^2)} .
\end{equation}

Using (\ref{equationmoments}), we get
%
%eC.2 #&#
%
\begin{equation}
\label{equationposteriormean}
\E(\beta_i \mid y_i) = \biggl\{ 1 - \frac{a'}{a' + b} \frac{\Phi
_1(b, 1, a' + b + 1, s', 1-1/\tau^2)}
{ \Phi_1(b, 1, a' + b, s', 1-1/\tau^2)} \biggr\} y .
\end{equation}
And by the law of total variance,
%
%eC.3 #&#
%
\begin{eqnarray}
\label{equationHBvariance}
\Var(\beta_i \mid y_i) &=& \E\{\Var(\beta_i \mid y_i, \kappa_i)
\} + \Var\{ \E(\beta_i \mid y_i, \kappa_i) \} \nonumber\\[-8pt]\\[-8pt]
&=& \sigma^2 \{1 - \E(\kappa_i \mid y_i)\} + y^2 \Var(\kappa_i
\mid y_i) \nonumber
\end{eqnarray}
with all other posterior moments for $\beta_i$ following in turn.

There is also a tractable expression for the marginal likelihood of the data:
%
%eC.4 #&#
%
\begin{equation}
\label{predictivewithintegral}
m(y_i) = C_1^{-1} \int_0^{1} \kappa_i^{a' - 1} (1-\kappa_i)^{b-1}
\biggl\{ \frac{1}{\tau^2} + \biggl(1 - \frac{1}{\tau^2} \biggr)
\kappa_i \biggr\}^{-1}
e^{-\kappa_i s'} \,\dd\kappa_i ,\hspace*{-30pt}
\end{equation}
where again $s' = s + y_i^2/(2 \sigma^2)$ and $a' = a + 1/2$. This
integral is in the same family as (\ref{hypergeometricintegral}), and
so by the same series of arguments we obtain
%
%eC.5 #&#
%
\begin{equation}
\label{generalpredictivey}\qquad
m(y_i) = \frac{1}{\sqrt{2\pi\sigma^2}} \exp\biggl( - \frac
{y_i^2}{2\sigma^2} \biggr) \frac{\Be(a', b)}{\Be(a, b)}
\frac{\Phi_1(b, 1, a' + b, s', 1-1/\tau^2)}
{\Phi_1(b, 1, a + b, s, 1-1/\tau^2)} .
\end{equation}
\end{appendix}

%
%We use the theory laid out in \cite{gordy1998} and
%The hypergeometric inverted-beta density is proper for all $a, b, \tau
%> 0$ and $s \in\mathbb{R}$.
%
%The normalizing constant in (\ref{generalhypbetaequation}) is
%C = \int_0^{1} \kappa^{\alpha- 1} (1 - \kappa)^{\beta- 1}
%Let $\eta= 1-\kappa$. Using the identity that $e^x = \sum_{m=0}^{
%$$
%C = e^{-s} \sum_{m=0}^{\infty} [ \frac{s^m}{m!} \int_0^1 \eta^{
%$$
%Using properties of the hypergeometric function $\phantom{}_2 F_1$
%some straightforward algebra,
%C = e^{-s} \ \Be(\alpha, \beta) \ \sum_{m=0}^{\infty} \sum_{n=0}^{
%{(\alpha+ \beta)_{m+n} \ m! \ n!}
%where $\Be(\cdot, \cdot)$ is the beta function and $(a)_n$ is the
%rising factorial. Appendix C of \cite{gordy1998} proves that, for all $
%yielding
%C = e^{-s} \ \Be(\alpha, \beta) \ \Phi_1(\beta, 1, \alpha+ \beta, s,
%1-1/\tau^2) ,
%where $\Phi_1$ is the degenerate hypergeometric function of two
%variables \citep[9.261]{gradshteyn:ryzhik:1965}.

\section*{Acknowledgments}

The authors would like to thank Mumtaz Ahmed
and Michael Raynor of Deloitte Consulting\vadjust{\goodbreak} for their insight into the
problem described here. They also acknowledge the helpful advice of two
anonymous referees. Finally, they thank Jake Benson of the University
of Texas for his help in collecting data on companies mentioned in
popular management books.\looseness=-1

%suskaldyti doi

% imsref loaded by lrinkeviciute, 2011-12-09 11:29:03
% imsref loaded by lrinkeviciute, 2011-12-09 12:33:56
%

\printaddresses

\end{document}